\documentclass[aps,superscriptaddress,showpacs,prl,twocolumn]{revtex4}
\usepackage{graphicx}
\usepackage{times}

\begin{document}

\title{Electronic Transport through Magnetic Molecules with Soft Vibrating Modes}
\author{P. S. Cornaglia}
\affiliation{Instituto Balseiro and Centro At\'omico Bariloche, Comisi\'on Nacional de Energ\'{\i}a At\'omica, 8400 San Carlos de Bariloche, Argentina.}
\affiliation{Centre de Physique Th\'eorique, \'Ecole Polytechnique, CNRS, 91128 Palaiseau Cedex, France.}
\author{Gonzalo Usaj}
\affiliation{Instituto Balseiro and Centro At\'omico Bariloche, Comisi\'on Nacional de Energ\'{\i}a At\'omica, 8400 San Carlos de Bariloche, Argentina.}
\author{C. A. Balseiro}
\affiliation{Instituto Balseiro and Centro At\'omico Bariloche, Comisi\'on Nacional de Energ\'{\i}a At\'omica, 8400 San Carlos de Bariloche, Argentina.}

\date{September 26, 2007}

\begin{abstract}
The low-temperature transport properties of a molecule
are studied in the field-effect transitor geometry.  
The molecule has an internal mechanical mode that modulates its 
electronic levels and renormalizes both the interactions and the coupling 
to the electrodes.
For a soft mechanical mode the spin fluctuations in the molecule are 
dominated by the bare couplings while the valence changes 
are determined by the dressed energies. 
In this case, the transport properties present an anomalous behavior
and the Kondo temperature has a weak gate voltage dependence.
These observations are in agreement with recent experimental data.
\end{abstract}

 

\maketitle
The recent development of molecular transistors (MT) has created new
scenarios for the study of correlation effects in nanoscopic systems.  
These devices have attracted a lot of interest due to their potential application
in nanoelectronics and their rich variety of behavior. 
Molecular transistors consist of a small molecule
connecting two electrodes and, in most cases, a gate electrode
is used to control the molecule's charge electrostatically. 
The transport properties of MT show signatures of strong electronic correlations
as Coulomb blockade
~\cite{JPark_2002} and the Kondo effect
\cite{WLiang_2002,AbhayN.Pasupathy10012004,Yu2004,Yu2005} 
similar to those observed in quantum dot devices (QD)~\cite{Gordon1998,Kouwenhoven2001}.
A remarkable difference between QD and MT is the coupling, on the latter, 
of the electronic degrees of freedom with a discrete set of mechanical modes~\cite{ParkPLAAM2000,Zhitenev2002,Qiu2004}. 
 
For the simplest case of a linear modulation of the molecule's 
electronic levels 
by a single vibration mode, several effects are predicted to occur.
The Franck-Condon renormalization of the molecule-electrodes
coupling is expected 
to produce a suppression in the sequential and cotunneling transport 
through the molecule~\cite{Flensberg_2003,vonOppen2005,Koch2006}.
The reduction of the effective Coulomb repulsion in the molecule may
lead to an effective \textit{e-e} \textit{attraction}
~\cite{Alexandrov2003a,cornaglia2004,Arrachea2005} and a strong 
sensitivity to gate voltage.

In the \textit{repulsive} \textit{e-e} interaction regime the spin-Kondo effect
dominates the low-temperature physics.
The Kondo effect generates in these devices an increase of 
the conductance with decreasing temperature and a zero-bias peak in the
differential conductance. 
These observations are a direct consequence of the formation of the Abrikosov-Suhl or Kondo
resonance below the Kondo temperature $T_{K}$ \cite{Hewson-book}.

In some MTs based on organometallic molecules, an \textit{anomalous} 
gate voltage dependence of the transport properties has been reported 
\cite{WLiang_2002,JPark_2002,Yu2005}. 
In the transition metal
complexes studied by Yu \textit{et al.} \cite{Yu2005}, $T_{K}$
depends weakly on the applied gate voltage and shows a rapid increase only
close to the charge degeneracy points. Moreover, the edges of the Coulomb
blockade diamonds are not well defined in the Kondo charge state. Such a 
behavior is inconsistent with the usual theory based on the Anderson model.

In this Rapid Communication we present a study of the Anderson-Holstein model showing
results obtained with the numerical renormalization group (NRG) \cite%
{BullaRMP,Hewson_2002}. We
find that, as the frequency of the vibrating mode decreases, an anomalous
gate dependence of $T_{K}$ and of the transport properties emerges.
 This effect arises
because the soft vibrating modes in the MT drive the system into a new
regime where the characteristic energy scales for spin and charge fluctuations
are not related as in the conventional theory of the Kondo effect.

The model Hamiltonian is $H=H_{M}+H_{E}+H_{ME}$ where the first two terms
describe the isolated molecule and the electrodes, respectively, and the last
term describes their coupling. We have 
\begin{eqnarray}
H_{M} &\!\!=\!\!&\varepsilon _{d}n_{d}\!+\!Un_{d\uparrow }n_{d\downarrow
}\!-\!\lambda \left( n_{d}\!-\!1\right) \left( a\!+\!a^{\dagger }\right)
\!+\!\omega _{0}a^{\dagger }a\,\!\qquad \\
H_{E} &\!=\!&\sum_{{k},\sigma ,\alpha }\varepsilon _{\alpha {k}}\;c_{\alpha {%
k}\sigma }^{\dagger }c_{\alpha {k}\sigma }^{}, \\
H_{ME} &\!=\!&\sum_{{k},\sigma ,\alpha }V_{\alpha k}( d_{\sigma
}^{\dagger }\;c_{\alpha {k}\sigma }\!+\!c_{\alpha {k}\sigma }^{\dagger
}d_{\sigma }^{}) .  \label{hamil}
\end{eqnarray}
Here $n_{d}={n}_{d\uparrow }+{n}_{d\downarrow }$, ${n}_{d\sigma }=d_{\sigma
}^{\dagger }d_{\sigma }$, $d_{\sigma }^{\dagger }$ creates an electron at
the molecular orbital with energy $\varepsilon _{d}$, $U$ is the
intramolecular Coulomb repulsion and $c_{\alpha k\sigma }^{\dagger }$
creates an electron in the mode $k$ of electrode $\alpha =L$, $R$. The
operator $a^{\dagger }$ creates an excitation of the vibronic mode with
energy $\omega _{0}$. The Fermi energy $E_F$ is set to zero, $\hbar =1$, and
all energies are in units of half the electrodes' bandwidth. For the
sake of simplicity, from here on we consider a symmetric molecule with
identical $L$ and $R$ electrodes ($\varepsilon _{\alpha k}\equiv \varepsilon
_{k}$) and take $V_{Lk}=V_{Rk}\equiv V_{hyb}$.

\begin{figure}[tbp]
\centering
\includegraphics[width=0.40\textwidth,clip]{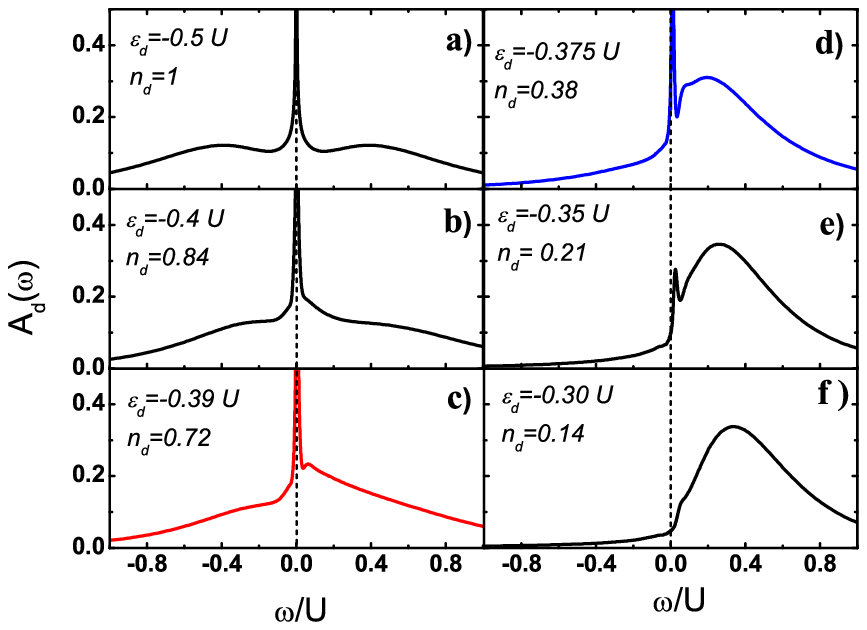} \centering
\includegraphics[width=0.40\textwidth,clip]{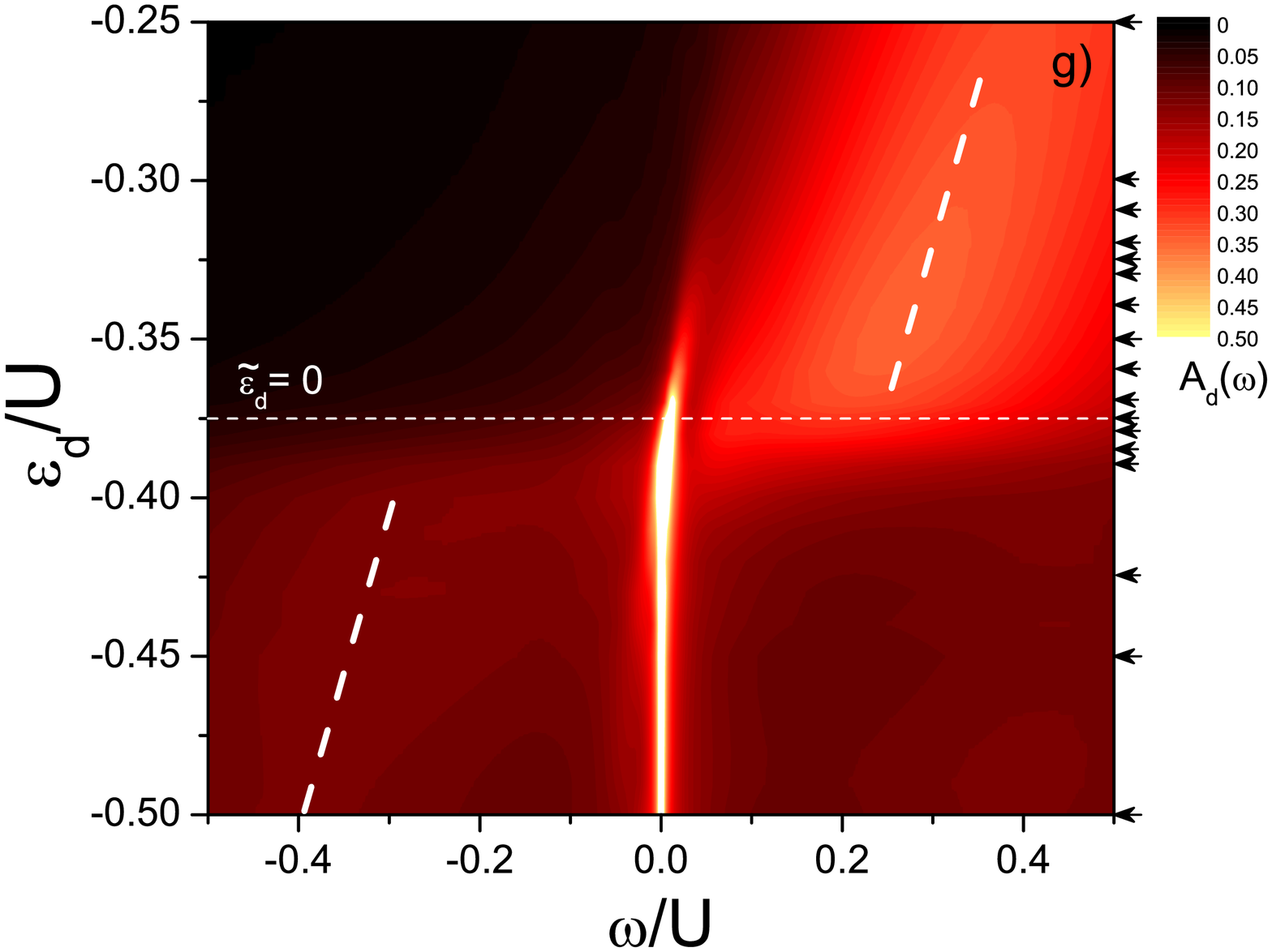}
\caption{(color online) Spectral density $A_{d}(\omega )$ (in units of $%
1/\pi\Gamma$). Panel a) to f) $A_{d}(\omega )$ vs $\omega $ for $U=0.2$, $%
V_{hyb}=0.1/\protect\sqrt{2}$, $g=6.25$, $\lambda=0.03$ and different values
of $\varepsilon _{d}$. g) Color map of $A_{d}(\omega )$ constructed by linear
extrapolation of the NRG results obtained for different values of $\varepsilon
_{d}$ (indicated by arrows). Thick dashed lines are a guide to the eye
of the form $\varepsilon _{d}=\omega +a$ and the horizontal line indicates the charge degeneracy point for the isolated molecule ($\widetilde{\varepsilon }_{d}=0$). Values of $A_d(\omega)$ larger
than $0.5$ are white in the color scale.}
\label{fig1}
\end{figure}

The energy spectrum of the isolated molecule ($V_{hyb}=0$) can be readily
obtained. The energies for the $n_{d}=0$, $1$ and $2$ charge states, with $m$
vibron excitations, are $E_{0,m}=-\lambda ^{2}/$ $\omega _{0}+m\omega _{0}$, 
$E_{1,m}=\varepsilon _{d}+m\omega _{0}$, and $E_{2,m}=-\lambda ^{2}/\omega
_{0}+2\varepsilon _{d}+U+m\omega _{0}$, respectively. The corresponding
states are indicated as $\left\vert 0,m\right\rangle $, $\left\vert \sigma
,m\right\rangle $, and $\left\vert 2,m\right\rangle $ where $\sigma $ is the
spin index in the $n_{d}=1$ charge sector---note that the vibronic states for
different charge states correspond to different equilibrium positions of the
coordinate associated with the vibronic motion. It is convenient to define
the effective single-electron energy and \textit{e-e} repulsion as $%
\widetilde{\varepsilon }_{d}=E_{1,0}-E_{0,0}=\varepsilon _{d}+\lambda
^{2}/\omega _{0}$ and $\widetilde{U}=E_{2,0}+E_{0,0}-2E_{1,0}=U-2\lambda
^{2}/\omega _{0}$, respectively. The charge degeneracy points of the
isolated molecule are given by $\widetilde{\varepsilon }_{d}=0$ and $%
\widetilde{\varepsilon }_{d}+\widetilde{U}=0$. When the molecule is coupled
to the electrodes, the charge fluctuations are controlled by the Franck-Condon (FC) 
factors $\gamma _{n,m}=\left\vert \left\langle \sigma ,n\left\vert d_{\sigma
}^{\dagger }\right\vert 0,m\right\rangle \right\vert ^{2}$. 
We define $\gamma_{m}\equiv\gamma_{0,m}=e^{-g}g^{m}/m!$,
where $g=(\lambda/\omega_0)^2$ is a dimensionless parameter. For $g\gg 1$,
transitions between low-lying states are exponentially suppressed and the
system is in the FC \textit{blockade} regime \cite{Flensberg_2003,vonOppen2005}.

We now present numerical results for the molecular spectral density $A_{d}(\omega)$ 
which determines the transport properties of the MT. $A_{d}(\omega )$ is given 
by $-(1/\pi)\mathrm{Im}G_{dd}(\omega )$, where $%
G_{dd}(\omega )$ is the electronic Green's function of the molecule in the
presence of the electrodes. The zero-temperature NRG results for $%
A_{d}(\omega )$ are shown in Fig. \ref{fig1} for different values of the
parameters in a regime of strong FC effect ($g=6.25$). In the electron-hole
symmetric case [$\varepsilon _{d}=-U/2$, Fig. \ref{fig1}a)], $A_{d}(\omega )$
shows broad structures at the bare energies $\varepsilon _{d}$ and $%
\varepsilon _{d}+U$ and a well defined Kondo peak at the Fermi level.
Although not well resolved in the figure, peaks are also obtained at $%
\widetilde{\varepsilon }_{d}$ and $\widetilde{\varepsilon }_{d}+\widetilde{U}
$ ---these are the first FC sidebands. The evolution of the spectral density 
$A_{d}(\omega )$ as the molecular energy is shifted upwards presents
interesting features. First, the broad peaks and the first FC peaks--which are
better resolved for small $\left\vert \widetilde{\varepsilon }%
_{d}\right\vert $ [see Fig. \ref{fig1}c)]-- shift with the molecular energy,
while the Kondo temperature, as given by the width of the Kondo peak, is not
very sensitive to it. As $\widetilde{\varepsilon }_{d}$ crosses $E_{F}$, the
behavior of $A_{d}(\omega)$ indicates a rapid change in the occupation of
the molecular orbital $\left\langle n_{d}\right\rangle$. There is no 
crossing of a wide resonance through the Fermi level, as it occurs in the $g=0$
case. Here instead, there is a rapid transfer of spectral weight from 
$\omega\sim\varepsilon_d$ to $\omega\sim\varepsilon_d+U$,
in agreement with
the \textit{atomic limit} results \cite{Hewson_2002}. The change in
the structure of $A_{d}(\omega )$ occurs as the Kondo peak evolves into a FC
peak at small $\omega$. This is better seen in the color map of the
spectral density shown in Fig. \ref{fig1}g). This anomalous behavior has
important consequences on the thermodynamic and transport properties of the
system.

\begin{figure}[tbp]
\centering
\includegraphics[height=3.1cm,clip]{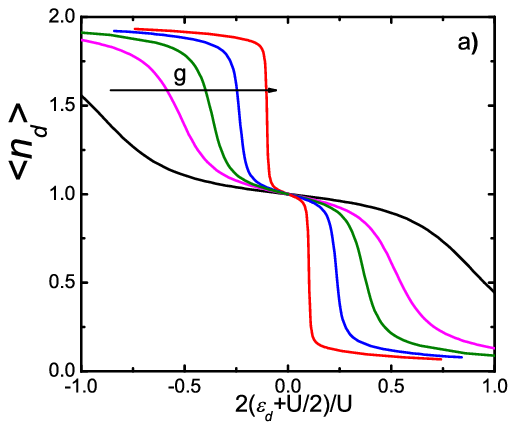} %
\includegraphics[height=3.1cm,clip]{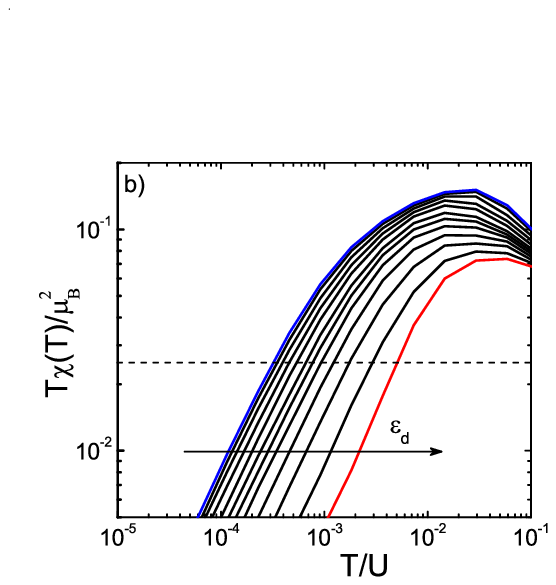} \centering
\includegraphics[width=0.33\textwidth,clip]{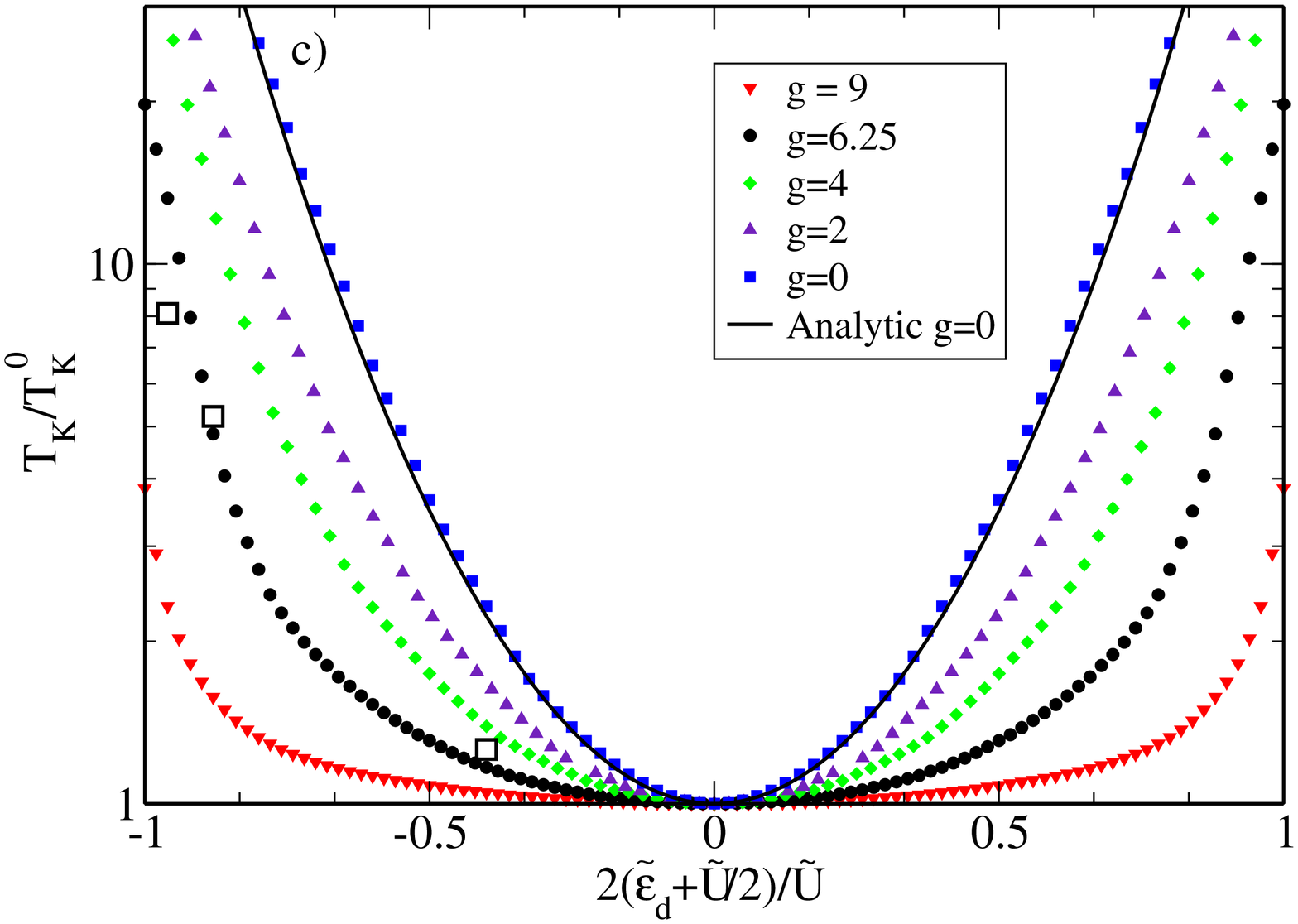}
\caption{(color online) a) Occupation of the molecular orbital $\left\langle
n_{d}\right\rangle $ vs $\varepsilon _{d}$ for $g=0$, $2$, $4$, $6.25$ and $%
9 $. b) Susceptibility $T\chi (T)/\mu _{B}{}^{2}$ vs $T$ \ in a log-log plot
for $\varepsilon _{d}/U$ increasing from $-0.5$ to $-0.38$, and $g=6.25$.
Dashed line indicates Wilson's value. Other parameters as in Fig. \ref{fig1}%
. c) Kondo temperature $T_K/T_K^{0}$ vs $\varepsilon _{d}$ for different
values of $g$ as indicated in the inset. Open squares are the values of $%
T_{K}$ obtained by fitting the temperature dependence of the conductance for 
$g\!=\!6.25$ (see text).}
\label{fig2}
\end{figure}

The occupation of the molecular orbital versus the bare energy $\varepsilon
_{d}$ is shown in Fig. \ref{fig2}a) for different values of $g$. As $g$
increases, the width of the region where the magnetic configuration is
stable ($\left\langle n_{d}\right\rangle \simeq 1$) and the crossover widths
between the different charge states decrease. The former is due to the
reduction of the effective \textit{e-e }repulsion $\widetilde{U}$ while the
latter, being exponential with $g$, is a manifestation of FC effects 
\cite{Flensberg_2003,vonOppen2005}. Similar FC effects are obtained
in the spinless case~\cite{Flensberg_2003}. 
However, for the Anderson-Holstein
model presented here, the interplay between \textit{e-e} and \textit{e-v}
interactions leads to a crossover between different charge states that is 
wider than in the spinless case. Also, for the spinfull model
with large $g$ the charge degeneracy points (with $\left\langle
n_{d}\right\rangle =1/2$ and $3/2$) are not given exactly by $\widetilde{%
\varepsilon }_{d}=0$ and $\widetilde{\varepsilon }_{d}+\widetilde{U}=0$ as
self-energy corrections, due to the hybridization, shift the dressed
molecular levels. 

The magnetic susceptibility $\chi (T)$ of the molecule 
is shown in Fig.~\ref{fig2}b) for different
values of $\epsilon_d$. In the Kondo regime, the low-temperature
susceptibility curves collapse into a single curve
when the temperature is properly scaled.
To calculate the Kondo temperature $T_{K}$
and its dependence on the molecular energy $\varepsilon _{d}$ we use
Wilson's criterion \cite{Wilson1975}: $T_{K}\chi (T_{K})/\mu _{B}{}^{2}=0.025
$. The results for $T_K$ are shown in Fig. \ref{fig2}c). As $\omega_0$ decreases 
(we keep $\lambda$ fixed and therefore $g$ increases), $T_{K}$ vs. 
$\widetilde{\varepsilon }_{d}$ deviates from the usual
behavior: it shows a weak dependence on $\widetilde{\varepsilon }_{d}$ at
the center of the Coulomb blockade region and a fast increase close to the
charge degeneracy points. 

In terms of the Kondo coupling $J$ the Kondo temperature is given by $%
T_{K}=De^{-1/\rho _{0}J}$ where $D$ is a high energy cutoff and $\rho _{0}$
is the bare electronic density of states of the electrodes at the Fermi
level. While the value of $J$ obtained using second order perturbation 
theory in $V_{hyb}$ underestimates the ratio  $%
T_{K}/T_{K}^{0}$ for small values of $\omega_0$, it is instructive to use 
this approach for a qualitative
interpretation of the numerical results. We have
\cite{balseiro:235409,Hotta2007} 
\begin{equation}
J=\sum_{m=0}\left( \frac{2V_{hyb}^{2}\gamma _{m}}{-\widetilde{\varepsilon }%
_{d}+\omega _{0}m}+\frac{2V_{hyb}^{2}\gamma _{m}}{\widetilde{\varepsilon }%
_{d}+\widetilde{U}+\omega _{0}m}\right) \,.  \label{eq:JS}
\end{equation}
For large $g$, the FC factor $\gamma _{m}$ 
is peaked at $m\sim m^{\star }=g$ with a width of the order of $\sqrt{%
m^{\star }}$. As in the summation of Eq.~(\ref{eq:JS}) the denominators are
slowly varying around $m^{*}$, the Kondo coupling can be approximated as $%
J\simeq 2V_{hyb}^{2}[1/(-\widetilde{\varepsilon }_{d}+m^{\star }\omega
_{0})+1/(\widetilde{U}+\widetilde{\varepsilon }_{d}+m^{\star }\omega _{0})]$%
. This shows that the relevant virtual charge fluctuations have a
characteristic energy given by the bare molecular parameters $\varepsilon
_{d}=\widetilde{\varepsilon }_{d}-\omega _{0}m^{*}$ and $\varepsilon _{d}+U=%
\widetilde{\varepsilon }_{d}+\widetilde{U}+\omega _{0}m^{*}$ corresponding
to the broad peaks in Fig. \ref{fig1}a). However, as $\widetilde{\varepsilon 
}_{d}\rightarrow 0$ $(-\widetilde{U})$ the $m=0$ term in Eq.~(\ref{eq:JS})
diverges with an exponentially small prefactor $e^{-g}$, indicating that the
perturbation theory breaks down for $\widetilde{\varepsilon }_{d}$
exponentially close to the charge degeneracy points of the isolated molecule.
In other words, for small $\omega_0$ and large $g$, the virtual charge 
fluctuations leading to the Kondo coupling are antiadiabatic while the 
charge instabilities are controlled by the dressed energies $\widetilde{\varepsilon }_{d}$
and $\widetilde{\varepsilon }_{d} + \widetilde{U}$.

\begin{figure}[tbp]
\centering
\includegraphics[width=0.34\textwidth,clip]{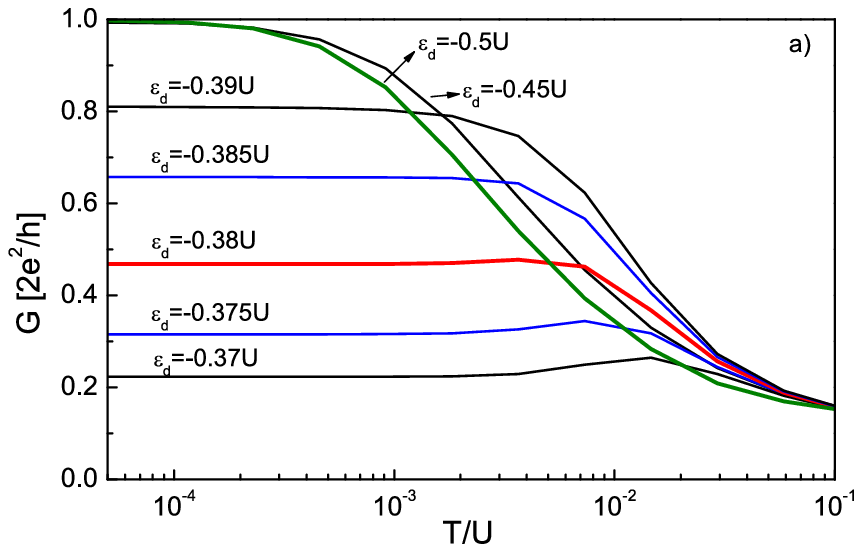} %
\includegraphics[height=3cm,clip]{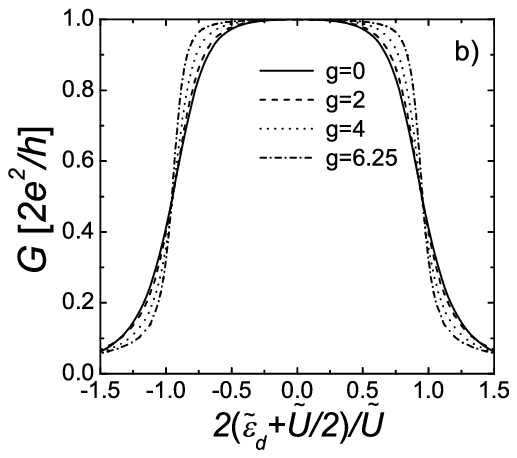} %
\includegraphics[height=3cm,clip]{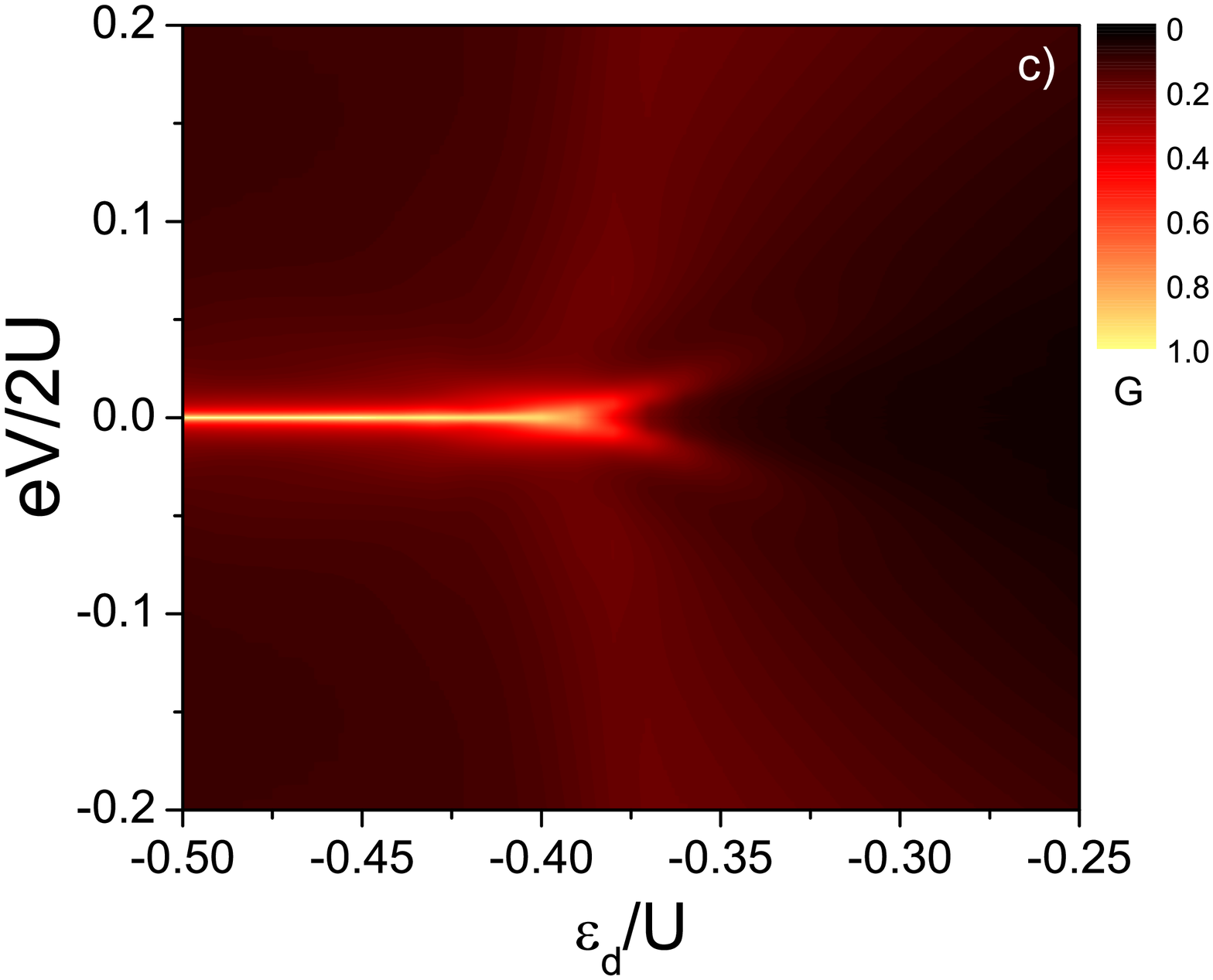}
\caption{(color online) a) Temperature dependence of the conductance for
different values of $\varepsilon _{d}/U$, and $g=6.25$. The charge
degeneracy point is at $\varepsilon _{d}/U\simeq-0.38U$. b) Zero-temperature 
conductance vs $\varepsilon _{d}$. Parameters as in Fig. \ref{fig1} 
and different values of $g$ as shown in the inset. c) Differential
conductance $G(V)$ (in units of $2e^2/h$) as a function of the bias and gate
voltages (cf. Eq. (\ref{gv})).}
\label{fig3}
\end{figure}

We stress that, even far from the degeneracy points, the perturbative approach of Eq.~(\ref{eq:JS}) gives a
quantitative estimate of $T_{K}/T_{K}^{0}$ only for large values of $\omega _{0}\gtrsim U/2$%
. As $\omega _{0}$ decreases the states $|\sigma
,m\rangle $ with $m>0$ also participate in the ground state and should be
taken into account \cite{PaaskeF05}. We are not aware of any reliable method to properly handle such a case analytically.

The linear conductance $G$ of the molecular junction at zero bias is given
by the spectral density $A_{d}(\omega )$ presented above. For symmetric
electrodes \cite{Pastawski1992,Meir1992} 
\begin{equation}  \label{eq:conductance}
G=\left. \frac{dI}{dV}\right| _{V=0}=\frac{e^{2}}{\hbar}\ \Gamma
\int_{-\infty }^{\infty }d\omega \left( -\frac{\partial f(\omega )}{\partial
\omega }\right) A_{d}({\omega })\;,
\end{equation}
where $f(\omega )$ is the Fermi distribution and $\Gamma =2\pi \rho
_{0}V_{hyb}^{2}$. The temperature dependence of the conductance for $g=6.25$
and different values of $\varepsilon _{d}$ is presented in Fig. \ref{fig3}%
a). In the electron-hole symmetric case ($\varepsilon_d=-U/2$) the
characteristic Kondo behavior is obtained. As $\varepsilon _{d}$ increases,
the low temperature conductance decreases in quantitative agreement with the
Fermi liquid zero-temperature results (see below). For $\widetilde{%
\varepsilon }_{d}>0$ the conductance shows a maximum at a temperature $T^{*}$
consistent with the energy of the first FC side-band. Within the Kondo regime our
results for the temperature dependent conductance are well fitted by the
phenomenological expression $G(T)=G_{0}^{*}[1+(2^{1/s}-1)(T/T_{K})^{2}]^{-s}$
with $s$ between $0.22$ and $0.25$ \cite{Gordon1998}. The value of $T_K$
obtained with this fitting procedure is consistent with Wilson's criterion
as shown in Fig. \ref{fig2}c). It is important to point out that while for $%
g=0$ the phenomenological expression correctly describes the universal
scaling of the conductance even deep into the intermediate valence regime ($%
\langle n_d\rangle\sim 0.5$), it fails to do so for $g\gg 1$ already for $%
\langle n_d\rangle\simeq0.75$ ($\varepsilon_d\simeq -0.39U$ for the
parameters of Fig. \ref{fig3}a).

The zero-temperature conductance, shown in Fig.~\ref{fig3}b) 
as a function of $\widetilde{\varepsilon}_d$ for different values of $g$, was 
calculated using 
\footnote{This expression is a consequence of Luttinger's theorem and has been shown 
to be exact for $g=0$ in the wideband limit~\cite{PhysRev.150.516,Hewson-book}. 
For $g\neq0$ it was verified numerically and a computation of a few low-order
diagrams in the expansion of the self-energy suggests that Luttinger's
theorem and Eq.~\ref{eq:luttG} can be generalized to the present 
situation~\cite{Cornaglia2005a}.} 
\begin{equation}  \label{eq:luttG}
G=\frac{e^2}{\hbar}\Gamma A_{d}(0)=\frac{2 e^2}{h}\sin ^{2}\left(\frac{\pi
\left\langle n_{d}\right\rangle}{2}\right).
\end{equation}
Note that while at high temperatures ($T>T_K$) the conductance is suppressed
by the Coulomb interaction and Franck-Condon blockade~\cite{Koch2006},
at low temperatures the Kondo effect sets in and the conductance is high.

A salient feature in the differential conductance of systems with strong \textit{e-v }
coupling is the anomalous behavior of the edges of the Coulomb blockade
diamonds. In some MT \cite{WLiang_2002,Yu2005} the diamond edges associated
with the Kondo charge state ($\langle n_d\rangle\simeq1$) are much weaker,
or even absent, than the ones corresponding to the non-Kondo charge state.
In the \textit{lowest order} on the bias voltage, the differential conductance is 
\begin{equation}
G(V)\simeq\frac{e^2}{h}\pi\Gamma[A_d(eV/2)\!+\!A_d(-eV/2)]\,.
\label{gv}
\end{equation}
Although the results obtained with this equation [see Fig.~\ref{fig3}c)] 
should only be taken as a rough estimate, they show that the qualitative behavior of the
differential conductance is very different from that of the Anderson model.
The main reason is that, as discussed above, the valence change is not due
to a broad resonant level that crosses $E_F$. The spectral weight is rather
transferred from below to above $E_F$ and only the narrow peak of the first
FC band, that has an exponentially reduced weight ($\propto e^{-g}$),
crosses $E_F$.

In summary, we have studied a model MT where the electronic levels of the
molecule are modulated by a well defined internal vibronic mode. 
We have shown that the coupling to a soft phonon mode ($\omega_0\ll U$) 
in the FC regime ($g\gg 1$)
changes qualitatively the behavior of the MT physical properties. The electronic
interactions in the molecule and the coupling to the electrodes are strongly renormalized.
While spin fluctuations in the molecule are associated to virtual 
processes dominated by the bare electronic energies, the valence changes occur when the dressed
energies cross the Fermi level. This leads to an anomalous gate voltage dependence of
the spectral density and $T_K$. The theory accounts for the observed weak
dependence of $T_{K}$ on gate voltage and suggests the possible origin of
the anomalous behavior of the Coulomb blockade diamond edges \cite{Yu2005}.
We found that the universality characteristic of the Kondo
phenomena, is lost in this case much before the valence change as compared 
with the usual Anderson model.
Finally, we showed that the FC suppression of the low-bias conductance 
expected for the high temperature regime~\cite{vonOppen2005,Koch2006} is not present 
at low temperatures due to the Kondo enhancement 
of the conductance.

This work was supported by ANPCyT Grants No 13829 and 13476 and CONICET PIP
5254. GU and PSC are members of CONICET (Argentina).

\end{document}